\documentclass[aps,prc,twocolumn,amsmath,superscriptaddress,floatfix,nofootinbib]{revtex4-2}

\usepackage{amsmath,amssymb,amsfonts,mathtools,bm}
\usepackage{graphicx}
\usepackage{subcaption}
\usepackage{xcolor,ulem}
\usepackage{braket}
\usepackage{hyperref}
\hypersetup{
  bookmarksnumbered=true,bookmarksopen=true,colorlinks=true,
  citecolor=blue,linkcolor=red,urlcolor=blue
}

\newcommand{\tred}{\textcolor{black}}
\newcommand{\ttred}{\textcolor{black}}

\begin{document}

\title{Probing hot QCD medium with heavy quarkonium in small and large collision systems}

\author{Jiamin Liu}
\affiliation{Department of Physics, Tianjin University, Tianjin 300350, China}

\author{Baoyi Chen}
\email[]{baoyi.chen@tju.edu.cn}
\affiliation{Department of Physics, Tianjin University, Tianjin 300350, China}
\affiliation{The International Joint Institute of Tianjin University,
Fuzhou, Tianjin University, Tianjin 300072, China}


\begin{abstract}
\tred{The yield ratios of different heavy-quarkonium states serve as sensitive probes of final-state interactions in relativistic nuclear collisions, because common cold-nuclear-matter effects are expected to be substantially reduced in these ratios.} To quantify hot QCD medium effects in small collision systems, such as proton-nucleus collisions, we employ a time-dependent Schr\"odinger equation framework to consistently simulate the real-time evolution of both bottomonium and charmonium states in the presence of in-medium complex heavy-quark potentials. In $p$-Pb collisions at $\sqrt{s_{\rm NN}}=8.16$ TeV, our model successfully describes the observed suppression in the yield ratios of excited-to-ground states—specifically $\Upsilon(nS)/\Upsilon(1S)$ and $\psi(2S)/J/\psi$—as a function of charged-particle multiplicity. This agreement supports the formation of a transient, hot QCD medium in small systems. Furthermore, the framework is employed to study the ratio of bottomonium nuclear modification factors in \(\sqrt{s_{\rm NN}}=5.02~\mathrm{TeV}\) Pb-Pb collisions, where hot medium effects become stronger. 
By establishing a unified description across two distinct heavy-quark flavors and different collision systems, our study indicates that the yield ratio of bottomonium states serves as a comparatively clean probe of the hot QCD medium generated in small collision systems.

\end{abstract}

\maketitle


Heavy quarkonia are among the most sensitive probes of the hot quantum chromodynamics (QCD) matter produced in relativistic nuclear collisions \cite{Matsui:1986dk,Andronic:2015wma}. 
\tred{ Over the past decades, the final yields and momentum distributions of heavy quarkonia have been extensively studied to constrain the initial energy density and transport properties of the bulk medium generated in nucleus-nucleus collisions~\cite{Rapp:2008tf,Emerick:2011xu,Du:2017qkv,Yan:2006ve,Chen:2016dke,Chen:2015iga,Zhao:2017yan,Zhou:2014kka,Andronic:2003zv}.}
When heavy quarkonia are in the hot deconfined medium, thermal light partons can partially screen the heavy quark potential inside quarkonium~\cite{Satz:2005hx}. Meanwhile, the parton inelastic scatterings from thermal partons also dissociate the quarkonium states~\cite{Laine:2006ns,Burnier:2007qm,Akamatsu:2020ypb}. Those hot-medium effects are closely connected with the final observables of heavy quarkonia and can be encoded in the complex heavy-quark potential~\cite{Rothkopf:2011db,Burnier:2014ssa,Margotta:2011ta,Krouppa:2017jlg,Wen:2022utn,Wen:2022yjx,Brambilla:2016wgg,Brambilla:2020qwo}.
As the binding energies of the heavy quarkonium states become different from each other, they can be regarded as a thermometer of the hot deconfined medium~\cite{Digal:2001ue,Wen:2022yjx}, where excited quarkonium states suffer stronger dissociation from the hot deconfined medium compared with the ground state~\cite{Brambilla:2010cs,Strickland:2011aa}. This is also called the sequential suppression pattern which has been observed in bottomonium states~\cite{Digal:2001ue,Wen:2022yjx}.

While heavy quarkonium dynamics in nucleus-nucleus collisions is mainly controlled by the hot-deconfined medium effects \cite{ALICE:2014wnc,CMS:2023gfb}, the situation in small (such as proton-nucleus, high-multiplicity proton-proton) collision systems remains less settled~\cite{CMS:2025psi2SJpsi_pPb}. The light hadron spectrum in high-multiplicity $pp$ and $p$-Pb collisions exhibits collective-like features, which have been well explained with the hydrodynamic model by assuming a small hot deconfined medium generated with shorter lifetimes ~\cite{CMS:2010ifv,CMS:2012qk,ALICE:2012eyl,Nagle:2018nvi,Zhao:2017rgg,Zhao:2020wcd}.
As heavy quark and quarkonium are produced in the initial parton hard scatterings, experience the early stage of the hot medium, they are regarded as a relatively clean probe of the early information of the hot medium in nuclear collisions. However, besides the final state interactions with the bulk medium, initial cold nuclear matter effects including shadowing effect and the Cronin effect can also alter the final production of the heavy quarkonium~\cite{Arleo:2012rs,Ferreiro:2014bia,Du:2018wsj}. These cold nuclear matter effects are not negligible compared with the small hot medium effects in $p$-Pb collisions.

\tred{Since common initial-state nuclear effects are expected to affect the production of the ground and excited quarkonium states in a similar manner, their contributions are substantially reduced in yield ratios such as $\Upsilon(2S)/\Upsilon(1S)$, $\Upsilon(3S)/\Upsilon(1S)$, and $\psi(2S)/J/\psi$.
These ratios therefore provide a comparatively clean probe of state-dependent final-state interactions with the hot medium.} In this work, we investigate the yield ratios of both charmonium and bottomonium simultaneously in both small ($p$-Pb)~\cite{CMS:2020fae,CMS:2025psi2SJpsi_pPb} and large (Pb-Pb)~\cite{ALICE:2014wnc,CMS:2023gfb} collision systems, by solving the time-dependent Schr\"odinger equation~\cite{Wen:2022utn,Wen:2022yjx,Margotta:2011ta,Krouppa:2017jlg} with an extracted complex in-medium potential in nucleus-nucleus collisions at RHIC 200 GeV and LHC 2.76 TeV and 5.02 TeV~\cite{Liu:2026uav}. 
The temperature profiles of the hot medium in $p$-Pb and Pb-Pb collisions are generated by the hydrodynamic model respectively~\cite{Shen:2014vra,Singh:2021evv,Zhao:2020wcd}. 
Our theoretical studies simultaneously explain well the normalized yield ratios of charmonium and bottomonium in both $p$-Pb and Pb-Pb collisions, covering heavy quarkonium dynamics with different quark flavors and distinct bulk medium. This study therefore establishes a consistent description about quarkonium dynamics, based on which a small hot deconfined medium is believed to be generated in $p$-Pb collisions.


Attributed to the large mass of heavy quarks, the inner evolution of heavy quarkonium can be well described with the time-dependent Schr\"odinger equation model. The radial part of the equation is separated as (in natural units $\hbar=c=1$),
\begin{align}
i \frac{\partial}{\partial t}\psi(r,t)
=
\left(
-\frac{1}{2m_\mu}\frac{\partial^2}{\partial r^2}
+V(r,T)
+\frac{L(L+1)}{2m_\mu r^2}
\right)\psi(r,t),
\label{eq:sch}
\end{align}
with the definition of $\psi(r,t)=rR(r,t)$, where $R(r,t)$ is the radial part of the wave function. 
$L$ is the orbital angular momentum, and the reduced mass is $m_\mu=m_Q/2$.  For bottom quark we take $m_b=4.62~\mathrm{GeV}$, while for charm quark we use $m_c=1.25~\mathrm{GeV}$. The hot medium effects on heavy quarkonium \tred{are} implemented through a complex potential, $V(r,T)=V_R(r,T)+V_I(r,T)$. Regarding the real part of the in-medium heavy quark potential, the Bayesian analysis~\cite{Zheng:2025bdc} about the quarkonium spectrum suggests that the color screening effect is very small and the in-medium heavy quark potential is close to the case of vacuum Cornell potential. The deep neural networks systematically analyzing the bottomonium experimental data points $R_{AA}$s from RHIC 200 GeV Au-Au collisions to LHC (2.76 TeV, 5.02 TeV) Pb-Pb collisions also indicate a weak color screening effect and the parametrized $V_R(T,r)$ which can give best explanation of bottomonium $R_{AA}$ across various collision systems are close to the vacuum case. Those phenomenological studies are consistent with the recent lattice QCD calculations about $V_R$~\cite{Burnier:2016mxc,Liu:2026uav}.

Therefore, in the $p$-Pb collisions with small hot medium and Pb-Pb collisions with big hot medium, we take the real part of the in-medium heavy quark potential to be the vacuum Cornell potential, 
\begin{equation}
V_R(r,T)=\sigma r-\alpha/r,
\end{equation}
The parameter \(\alpha=\pi/12\) and \(\sigma=0.2~{\rm GeV}^2\) are fitted according to the masses of quarkonium states in vacuum~\cite{Satz:2005hx}.

\begin{figure}[t]
  \centering
\includegraphics[width=\linewidth]{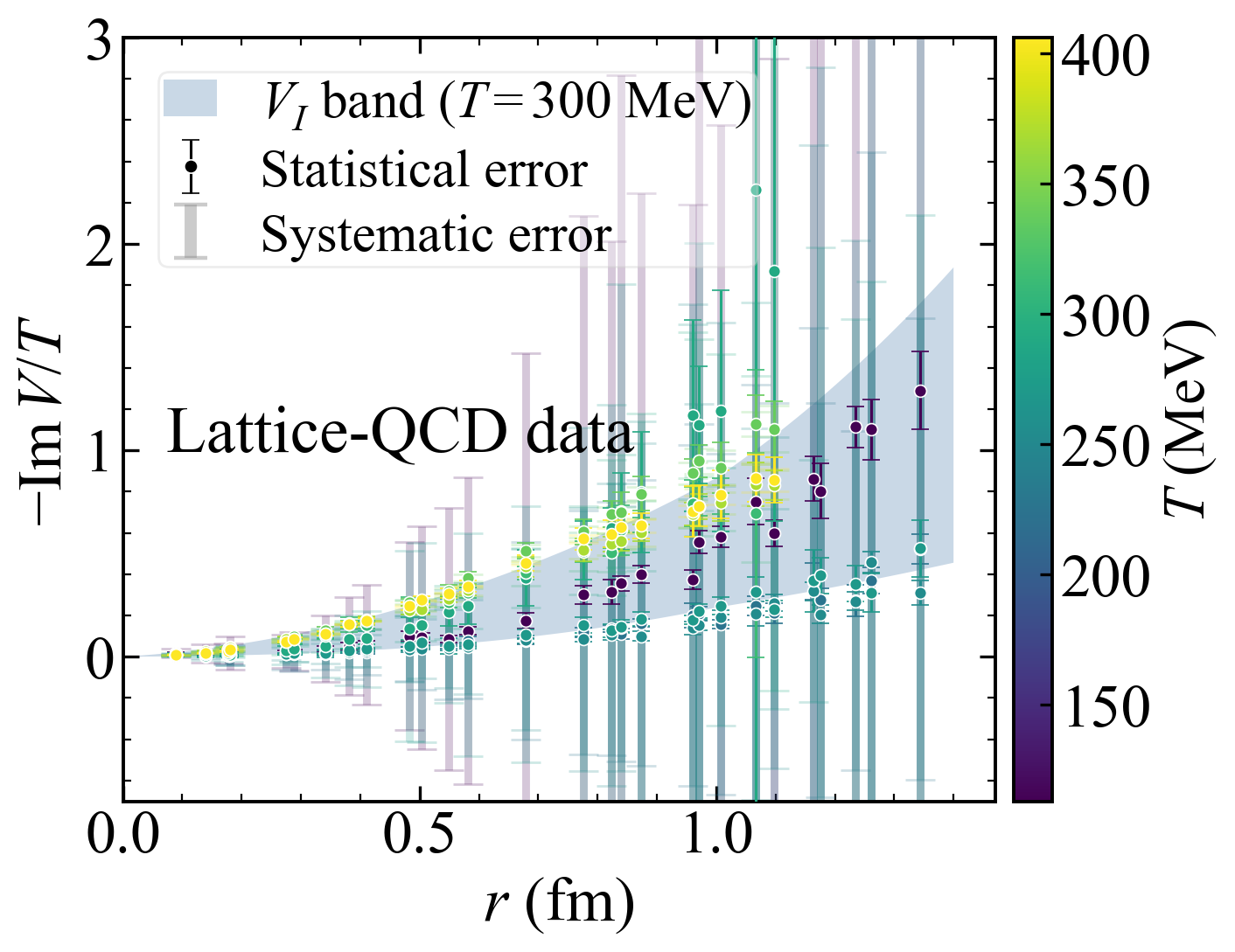}
\caption{
Dimensionless radial dependence of $-\mathrm{Im}\,V/T$ as a function of the interquark distance $r$ at $T=300~\mathrm{MeV}$. The colored points are lattice-QCD data from Ref.~\cite{Burnier:2016mxc}, with statistical and systematic uncertainties shown by black and gray error bars, respectively. \tred{The shaded band represents the uncertainty of \(V_I\) obtained from the previous deep neural network extraction~\cite{Liu:2026uav}, incorporating the experimental uncertainties of the fitted quarkonium data.}
}
\label{fig:VI}
\end{figure}

The random collisions from thermal partons are incorporated in the imaginary part of the potential $V_I$, which reduces the normalization of the color-singlet wave function $\psi(r,t)$,

\begin{equation}
\label{eq:VI}
V_I(r,T)=
-i\,{T}^{a_0}\Bigl(a_1\,r + a_2\,r^{\,a_3}\Bigr),
\end{equation}
Here, \(T\) is measured in GeV and \(r\) in fm, respectively. \tred{During the in-medium evolution, the wave function is not renormalized. The norm loss induced by the imaginary potential represents the depletion of the color-singlet component due to transitions out of the color-singlet sector. Only the color-singlet component is explicitly evolved in the present framework, while reverse transitions from color-octet to color-singlet states are neglected. Their impact on the ground-state bottomonium survival probability has been found to be small in previous studies~\cite{Brambilla:2022ynh,Wei:2026iwq}.} We incorporate the imaginary part of the heavy quark potential $V_I$ extracted by the deep neural networks previously~\cite{Liu:2026uav}, which can give the best explanation about bottomonium experimental data points $R_{AA}$ simultaneously. To account for the uncertainty of \(V_I\), we vary the parameters within the ranges  
$a_0\in[1.00,\,1.40]$, $a_1\in[0.034,\,0.18]$, $a_2\in[0.35,\,0.68]$, $a_3\in[2.00,\,2.57]$. The corresponding imaginary part of the potential at a fixed temperature \(T=0.3~{\rm GeV}\) is shown in Fig.~\ref{fig:VI}.

To compare with the experimental data, $Q\bar Q$ initial wave function is initialized with each quarkonium eigenstate. After evolving this wave function with the time-dependent Schr\"odinger equation in the hot medium, it is then projected to each eigenstate of quarkonium $\{\Phi_{nl}(r)\}$, i.e. $\psi(r,t)/r=\sum_{nl}c_{nl}(t)\Phi_{nl}(r)$, 
where the square of the coefficients $|c_{nl}(t)|^2$ represent the survival probability of one quarkonium moving out of the expanding hot medium. It depends on the local temperatures along the trajectory of the $Q\bar Q$ dipole. 
$n$ and $l$ \tred{are} the radial and angular quantum \tred{numbers} of each eigenstate. 

To evolve $Q\bar Q$ dipoles event-by-event in the heavy ion collisions, multiple dipoles will be generated with various random initial positions and the momentum. The \tred{position} of the heavy quark dipole at time $t$ is updated as $\mathbf{x}(t)=\mathbf{x}(t_0)+(t-t_0)\mathbf{p}/E$ with $E=\sqrt{m^2+|\mathbf{p}|^2}$~\cite{Wen:2022yjx}.

\tred{Here, \(t_0\) denotes both the hydrodynamic initialization time and the starting time of the Schr\"odinger evolution, with
\(t_0=0.2~\mathrm{fm}/c\) for \(p\)--Pb and \(t_0=0.6~\mathrm{fm}/c\) for Pb--Pb collisions. Pre-hydrodynamic evolution is not included. The radial Schr\"odinger equation is solved in the center-of-mass frame of the $Q\bar Q$ pair. At each time step, the local temperature obtained from the hydrodynamic medium is used as the input to the isotropic potential \(V(r,T)\).
The relative-velocity dependence between the quarkonium and the local fluid, and the associated anisotropic modification of the potential, are neglected in the present calculation~\cite{Liu:2012tn}.}
For the momentum $\mathbf{p}$ of $Q\bar Q$ dipole, it is sampled according to the distributions below, 
\begin{align}
\frac{dN}{2\pi p_T dp_T}
=
\frac{n-1}{\pi (n-2)\langle p_T^2\rangle_{pp}}
\left[
1+
\frac{p_T^2}{(n-2)\langle p_T^2\rangle_{pp}}
\right]^{-n},
\label{eq:pp-pT}
\end{align}
The parameters \(n\) and \(\langle p_T^2\rangle_{pp}\) are obtained by fitting the corresponding measured spectra of charmonium and bottomonium in pp collisions~\cite{Chen:2018kfo,Zhao:2021voa,Zhou:2014kka}. After evolving multiple $Q\bar Q$, the mean survival probability of a certain quarkonium eigenstate with the ensemble average can be written as, 
\begin{align}
\langle |c_{nl}(t)|^2 \rangle_{\rm en}=
\frac{
\int d\mathbf{x}\, d\mathbf{p}\,
|c_{nl}(t,\mathbf{x},\mathbf{p})|^{2}\,
\frac{dN}{d\mathbf{x}\,d\mathbf{p}}
}{
\int d\mathbf{x}\, d\mathbf{p}\,
\frac{dN}{d\mathbf{x}\,d\mathbf{p}}
},
\label{eq:enavg}
\end{align}
\tred{After including the feed-down contributions, the normalized prompt yields of the bottomonium states are given by}, 
\begin{align}
N\bigl(\Upsilon(kS)\bigr)
=
\sum_{n,l}
\left\langle |c_{nl}(t_f)|^{2}\right\rangle_{\rm en}
f_{\rm init}^{nl}
B_{nl\to kS},
\label{eq:yield_incl_Ups}
\end{align}
where \(k=1,2,3\), \(B_{nl\to kS}\) are the feed-down branching fractions in vacuum~\cite{Brambilla:2020qwo}, and \(f_{\rm init}^{nl}\) denote the weights of the different quarkonium initial production in the relevant kinematic acceptance. It is the product of the initial normalized transverse momentum distribution in Eq.~\eqref{eq:pp-pT} and the initial direct production cross section, which will be given below.


For charmonium observables, the CMS measurement reports the normalized multiplicity variable \(N_{\rm track}^{\rm corr}/\langle N_{\rm track}^{\rm corr}\rangle_{\rm MB}\), where tracks are counted within \(|\eta_{\rm track}|<2.4\) and \(p_T^{\rm track}>0.4~{\rm GeV}\), with \(\langle N_{\rm track}^{\rm corr}\rangle_{\rm MB}(|\eta_{\rm track}|<2.4)=59.5\pm0.5\)~\cite{CMS:2025psi2SJpsi_pPb}. To match the \(-1<y<1\) window used in the calculation, we estimate the corresponding mean multiplicity via the scaling, \(\langle N_{\rm track}\rangle_{\rm th}(|y|<1)\simeq24.8\).
The normalized yield ratio of different quarkonium states (labelled as $X$ and $Y$) is defined as,
\begin{align}
R_{X/Y}^{\rm norm}(i)
\equiv
\frac{
N^i(X)/N^i(Y)
}{
\left[\sum_j \omega_j N^j(X)\right]/
\left[\sum_j \omega_j N^j(Y)\right]
},
\label{eq:Rnorm_general}
\end{align}
where \(i\) and $j$ label the event-activity bins and the denominator defines the activity-integrated reference. We take \(\omega_j=N_{\rm coll}^j\) motivated by binary scaling of quarkonium production. \tred{Common cold-nuclear-matter effects, such as nuclear shadowing and Cronin broadening~\cite{Eskola:2016oht,Cronin:1974zm}, are expected to affect the production of \(X\) and \(Y\) in a similar manner and are therefore substantially reduced in their yield ratio. Motivated by this expected cancellation, cold-nuclear-matter effects are not explicitly included in the present calculation of the yield ratios. The ratios are consequently more sensitive to state-dependent final-state interactions with the hot medium.}
%


To study heavy quarkonium dynamics in the $p$-Pb and Pb-Pb collision systems, a proper description of the hot medium evolution is essential, which can be well described with the hydrodynamic models~\cite{Schenke:2010nt,Schenke:2010rr}. We employ the MUSIC package~\cite{Shen:2014vra} to obtain the temperature profiles of the hot QCD medium, which will be utilized in the Schr\"odinger equation for the heavy quarkonium evolution. The initial energy density profiles, which \tred{are inputs to} the hydrodynamic model, \tred{are constrained by} the final charged-particle multiplicity~\cite{Singh:2021evv}. In \(\sqrt{s_{NN}}=8.16~\mathrm{TeV}\) $p$-Pb collisions, the corresponding initial temperature can be obtained as, 
\begin{align}
T_0
=
\left[
\frac{90}{g_k\,4\pi^2}\,
C'\,
\frac{1}{A_T\tau_0}\,
1.5\,
\frac{dN_{\rm ch}}{dy}
\right]^{1/3},
\label{eq:T0_from_dNch}
\end{align}
where we utilize \(g_k=47.5\), \(C'\simeq3.6\), \(A_T=\pi R_T^2\) with \(R_T=0.9~{\rm fm}\), and \(\tau_0=0.2~{\rm fm}/c\) for the \(p\)--Pb
calculations. The value of \(\tau_0\), which defines the starting time of
the hydrodynamic evolution, is motivated by light-hadron observables in
small collision systems~\cite{Singh:2021evv,Zhao:2020wcd}. \tred{For the Pb--Pb calculations, we use \(\tau_0=0.6~{\rm fm}/c\)~\cite{Liu:2026uav}. Within each collision system, the hydrodynamic starting time is kept fixed across all event-activity or centrality classes, while the initial temperature or energy-density normalization is determined separately from the corresponding charged-particle multiplicity.} In $p$-Pb collisions, the track multiplicity is estimated as \(N_{\rm track}=\Delta y\,dN_{\rm ch}/dy\), where $\Delta y$ is the rapidity range of the experimental data. It is \(\Delta y=3.86\)  for the \(-1.93<y<1.93\) absolute-ratio comparison and \(\Delta y=4.79\)  for the \(-2.86<y_{\rm CM}<1.93\) normalized-ratio comparison.

\begin{figure}[!htp]
    \centering
\includegraphics[width=0.48\textwidth]{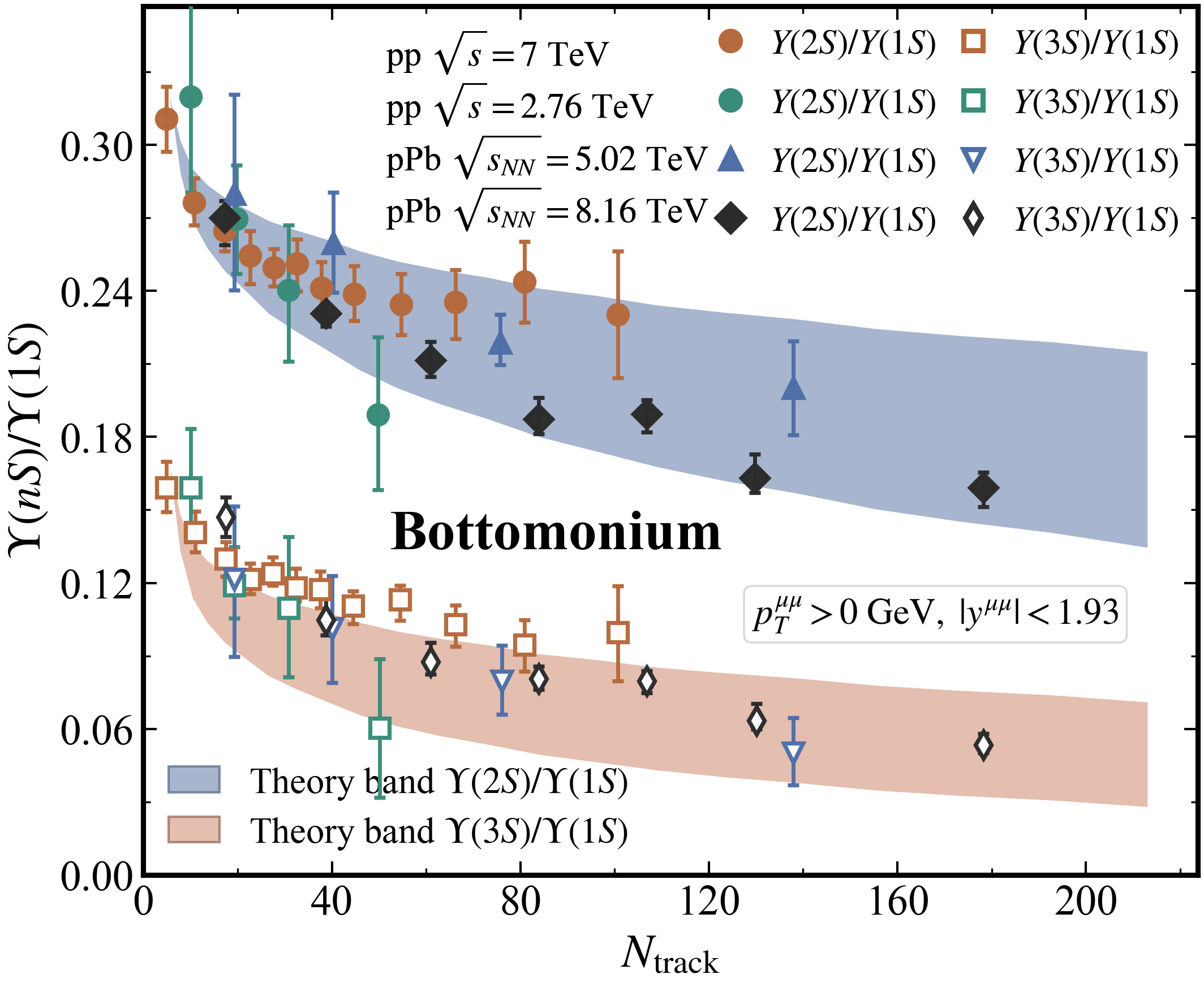}
\includegraphics[width=0.48\textwidth]{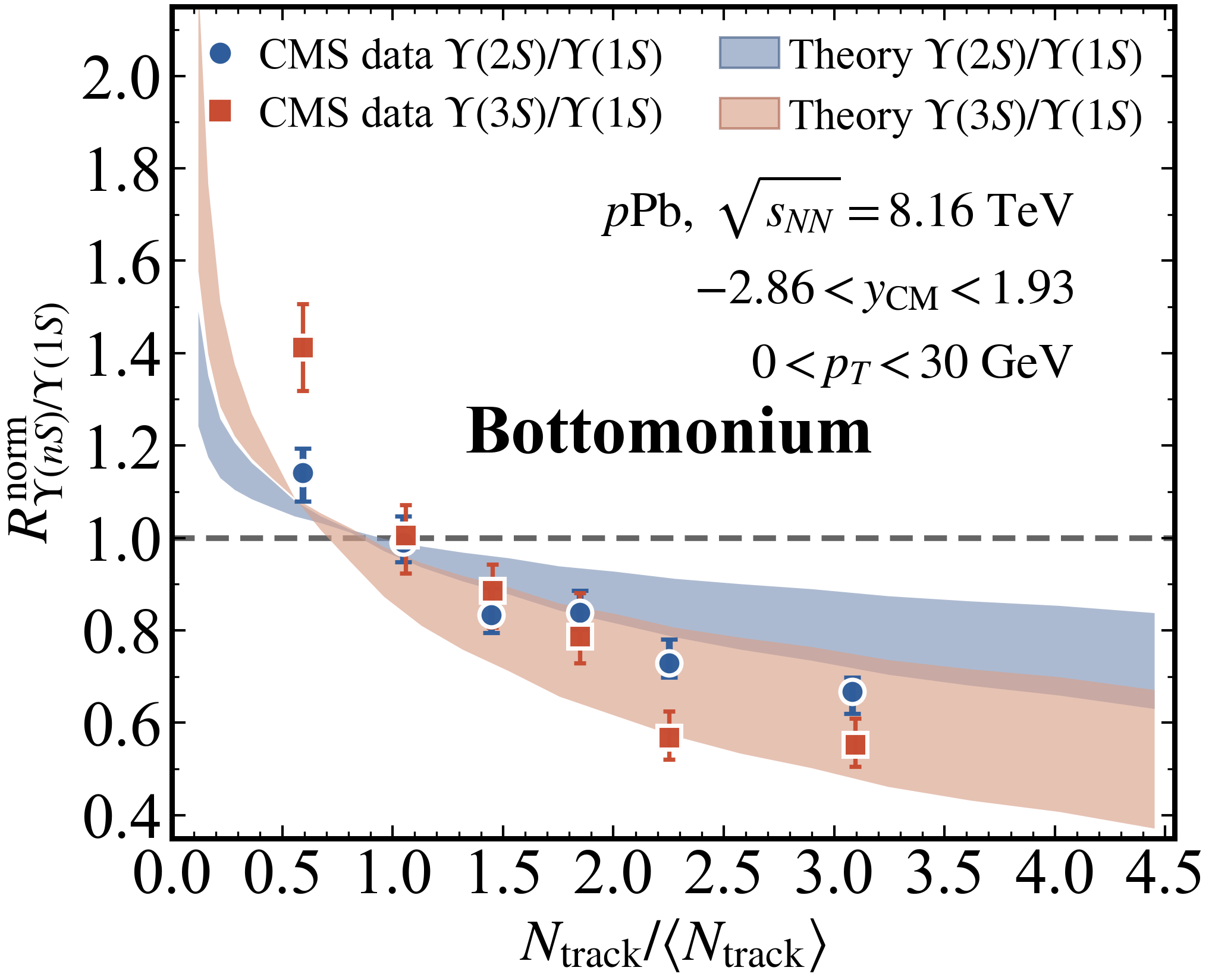}
    \caption{ Upper panel: Yield ratios of excited bottomonium states to the ground state, $\Upsilon(2\text{S})/\Upsilon(1\text{S})$ and $\Upsilon(3\text{S})/\Upsilon(1\text{S})$, as a function of $N_{\text{track}}$ in $p$-Pb collisions at $\sqrt{s_{\text{NN}}} = 8.16\text{ TeV}$. The results are compared with available $pp$ and $p$-Pb experimental data~\cite{CMS:2013dla,CMS:2020fae}. Lower panel: Normalized yield ratio $R^{\text{norm}}_{\Upsilon(2\text{S})/\Upsilon(1\text{S})}$ and $R^{\text{norm}}_{\Upsilon(3\text{S})/\Upsilon(1\text{S})}$ in $p$-Pb collisions at $\sqrt{s_{\text{NN}}} = 8.16\text{ TeV}$ as a function of $N_{\text{track}}/\langle N_{\text{track}}\rangle$. The shaded bands represent the uncertainties originating from the in-medium heavy-quark potential discussed in the text.
}
    \label{fig:ups_abs_norm}
\end{figure}


In Fig.~\ref{fig:ups_abs_norm}, the yield ratio of bottomonium $\Upsilon(2S)/\Upsilon(1S)$ and $\Upsilon(3S)/\Upsilon(1S)$ as a function of the $N_{\rm track}$ in  \(\sqrt{s_{NN}}=8.16~\mathrm{TeV}\) $p$-Pb collisions are calculated. Different from the ratio of quarkonium nuclear modification factors such as $R_{AA}(\Upsilon(2S))/R_{AA}(\Upsilon(1S))$ which mainly depends on the final interactions between quarkonium and the hot medium, the inclusive yield ratio $\Upsilon(2S)/\Upsilon(1S)$ \tred{is} closely connected with both hot medium effects and the initial production cross sections of each quarkonium state.

The bottomonium production inputs in Table~\ref{tab:pPb-cross-section} are constructed from an effective \(pp\) baseline at \(\sqrt{s_{NN}}=8.16~{\rm TeV}\) and unfolded with the vacuum feed-down matrix following Refs.~\cite{Brambilla:2020qwo,Liu:2026uav,Boyd:2023ybk}. The inclusive \(\Upsilon(1S)\) and \(\Upsilon(2S)\) cross sections are taken from the scaled \(pp\) reference of the LHCb \(p\)--Pb measurement~\cite{LHCb:2018psc}, and the \(\Upsilon(3S)\) input is obtained from the measured \(\Upsilon(3S)/\Upsilon(1S)\) ratio. The \(\chi_b(1P)\) and \(\chi_b(2P)\) components, which are not directly measured in the same kinematic region, are estimated by rescaling the \(\chi_b(nP)/\Upsilon(1S)\) ratios from the 5.02 TeV \(pp\) baseline~\cite{LHCb:2014ngh,Boyd:2023ybk}. The direct cross-section vector is determined from $\vec{\sigma}_{\rm dir}=F^{-1}\vec{\sigma}_{\rm incl}$, where \(F\) contains the effective vacuum branching fractions among bottomonium states~\cite{Brambilla:2020qwo}. These direct cross sections are used as the initial production weights \(f_{\rm init}^{nl}\) in Eq.~(\ref{eq:yield_incl_Ups}). At each $N_{\rm track}$ bin, the corresponding temperature profiles of the hot medium are utilized, which are given by the hydrodynamic model with the corresponding initial conditions. In the very small $N_{\rm track}$, the hot medium effect is expected to be negligible at $N_{\rm track}\rightarrow 0$, and the final yield ratio of $\Upsilon(2S)/\Upsilon(1S)$ and $\Upsilon(3S)/\Upsilon(1S)$ is the same as the initial condition.

\begin{table}[t]
\centering
\caption{Bottomonium production inputs used to determine \(f_{\rm init}^{nl}\) in the \(p\)--Pb calculation at \(\sqrt{s_{NN}}=8.16~{\rm TeV}\)~\cite{LHCb:2018psc,CMS:2013qur,CMS:2020fae,LHCb:2014ngh,Boyd:2023ybk}.
}
\label{tab:pPb-cross-section}
\begin{tabular}{c c c c c c}
\hline\hline
State
& \(\Upsilon(1S)\)
& \(\chi_b(1P)\)
& \(\Upsilon(2S)\)
& \(\chi_b(2P)\)
& \(\Upsilon(3S)\) \\
\hline
\(\sigma_{\rm incl}\) (nb) & 120.57 & 70.13 & 39.53 & 61.57 & 16.51 \\
\(\sigma_{\rm dir}\) (nb)  & 79.49  & 92.52 & 37.53 & 78.85 & 19.94 \\
\hline\hline
\end{tabular}
\end{table}

With increasing $N_{\rm track}$, the temperatures of the hot medium increase, which results in stronger suppression over the quarkonium wave function via the imaginary potential $V_I$. As the $V_I(T,r)$ becomes larger at larger radius $r$, the components of the excited quarkonium states in the heavy quark dipole are expected to be more reduced, attributed to their larger spatial size, compared with the ground state component. The bands of the theoretical calculations originate from the uncertainty of the parametrized $V_I$, which are determined via the deep neural networks by fitting the bottomonium experimental data points about $R_{AA}(N_p)$ and $R_{AA}(p_T)$ from RHIC 200 GeV Au-Au collisions, LHC 2.76 TeV and 5.02 TeV Pb-Pb collisions. In the absence of final-state hot-medium effects, the theoretical bands will be flat. The decreasing tendencies of $\Upsilon(2S)/\Upsilon(1S)$ and $\Upsilon(3S)/\Upsilon(1S)$ indicate that the bottomonium excited states are more affected by the hot medium compared with the ground state. Our theoretical results describe the experimental data from $p$-Pb collisions at $\sqrt{s_{NN}}=5.02$ TeV and 8.16 TeV and also pp collisions at $\sqrt{s_{NN}}=7$ TeV. Because the $\Upsilon(3S)$ component experiences stronger dissociation due to its larger spatial size, the predicted $\Upsilon(3S)/\Upsilon(1S)$ ratio is smaller than $\Upsilon(2S)/\Upsilon(1S)$.

To reduce the uncertainty of the $N_{\rm track}$ in experiments, the normalized ratio $R^{\rm norm}_{\Upsilon(nS)/\Upsilon(1S)}$, which is defined before, is presented as a function of $N_{\rm track}/\langle N_{\rm track}\rangle$, as shown in the lower panel of Fig.~\ref{fig:ups_abs_norm}. As the bottomonium $\Upsilon(3S)$ suffers stronger dissociation from the hot medium, therefore with increasing $N_{\rm track}/\langle N_{\rm track}\rangle$, the normalized yield ratio of $\Upsilon(3S)/\Upsilon(1S)$ shows stronger decreasing tendency compared with the case of the yield ratio of $\Upsilon(2S)/\Upsilon(1S)$, as shown in the lower panel of Fig.~\ref{fig:ups_abs_norm}. The theoretical bands qualitatively explain the experimental data points of the yield ratios $\Upsilon(nS)/\Upsilon(1S)$ in  \(\sqrt{s_{NN}}=8.16~\mathrm{TeV}\) $p$-Pb collisions.

\begin{figure}[htp]
    \centering
\includegraphics[width=\linewidth]{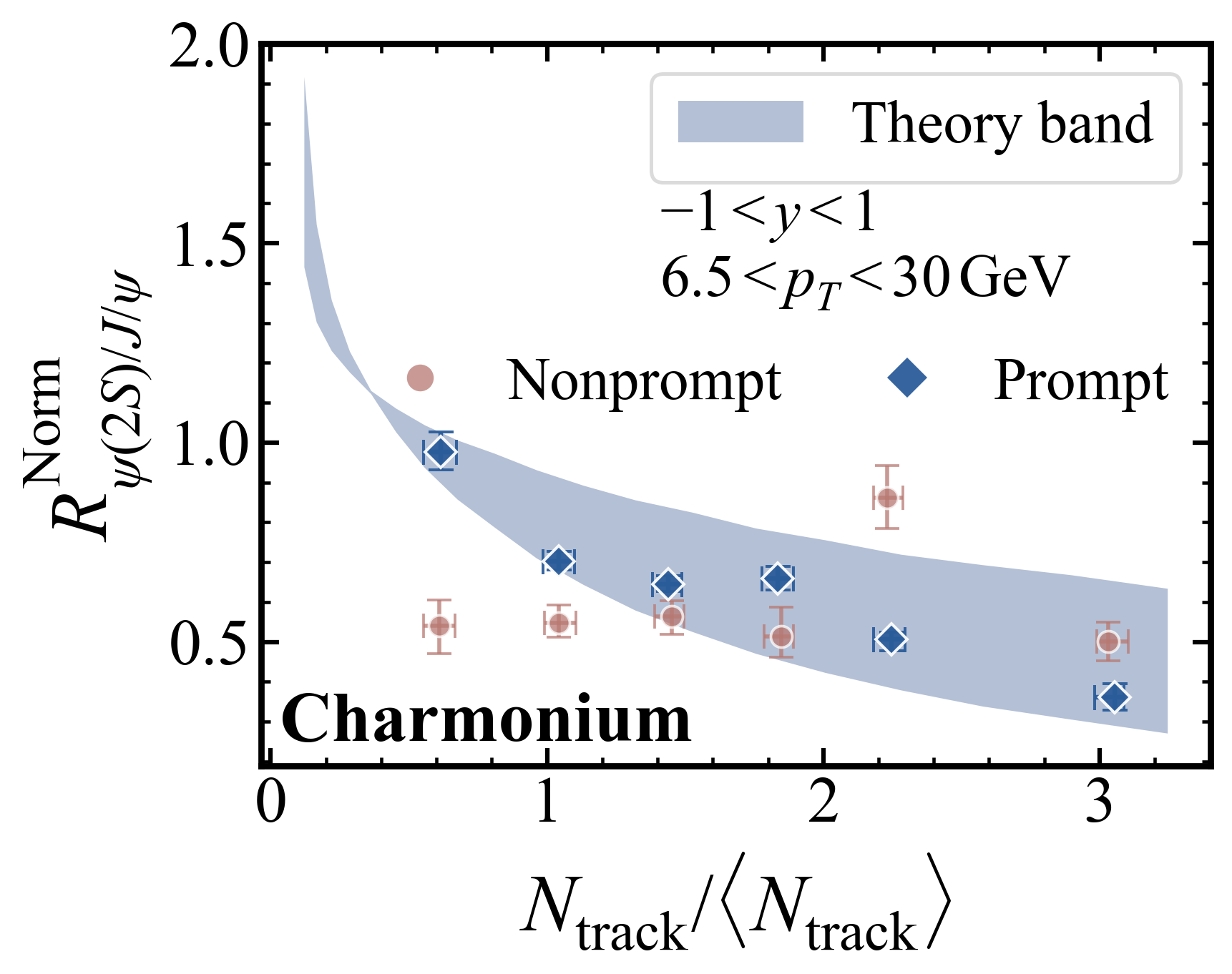}
    \caption{
    Normalized prompt yield ratios of $\psi(2S)$ over $J/\psi$ in \(p\)--Pb collisions at
    \(\sqrt{s_{NN}}=8.16~{\rm TeV}\) for \(-1<y<1\) and  \(6.5<p_T<30~{\rm GeV}/c\). The theoretical bands are for the prompt yield ratio, where the band originates from the uncertainty of the imaginary heavy quark potential $V_I$. The prompt and non-prompt experimental data points are cited from CMS Collaboration~\cite{CMS:2025psi2SJpsi_pPb}.
    }
    \label{fig:charmonium_norm}
\end{figure}

Based on the same Schr\"odinger equation model, we also study the normalized yield ratio of charmonium $\psi(2S)/J/\psi$ in  \(\sqrt{s_{NN}}=8.16~\mathrm{TeV}\) $p$-Pb collisions. For $J/\psi$, the prompt production of $J/\psi$ consists of the direct production and also the decay contribution from higher charmonium states (such as $\chi_c(1P)$ and $\psi(2S)$). According to the initial direct production cross sections of ($J/\psi$, $\chi_c(1P)$, $\psi(2S)$), we evolve those charmonium states in the hot medium created in $p$-Pb collisions, the prompt yield of $J/\psi$ and $\psi(2S)$ can be obtained after including the feed-down process. For the charmonium sector, their initial direct yield ratio of $(J/\psi, \chi_c, \psi(2S))$ is determined according to~\cite{Wen:2022utn}. The decay channels of $\chi_c\rightarrow J/\psi$ and $\psi(2S)\rightarrow J/\psi$ are considered, and the direct production cross section of three charmonium states without feed-down contributions are extracted to be $f_{pp}^{J/\psi}:f_{pp}^{\chi_c}:f_{pp}^{\psi(2S)}=0.68 : 1 : 0.19$~\cite{ParticleDataGroup:2018ovx, Wen:2022utn}.

In Fig.~\ref{fig:charmonium_norm}, our theoretical results of the normalized yield ratio $R^{\rm norm}_{\psi(2S)/J/\psi}$ decreases significantly with the $N_{\rm track}/\langle N_{\rm track}\rangle$, which indicates that the $\psi(2S)$ suffers stronger suppression in hotter medium compared with the ground state $J/\psi$. Our theoretical results explain well the prompt experimental data points. While for the non-prompt charmonium, they are from the decay of the B-hadron after bottom quark suffer energy loss in the hot medium and hadronize into B-hadrons. Different from charmonium bound states which can be dissociated in the medium, bottom quarks only suffer energy loss in the hot medium which happen before the hadronization of $b\rightarrow B$ and the feed-down process of $B\rightarrow J/\psi, \psi(2S)$. Therefore, hot medium effects encoded in the final distribution of bottom quarks \tred{contribute almost equally to} the non-prompt $J/\psi$ and $\psi(2S)$ from the B-hadron decay. Therefore, the non-prompt experimental data points show flat tendency with $N_{\rm track}/\langle N_{\rm track}\rangle$. 

\begin{figure}[htp]
    \centering
\includegraphics[width=\linewidth]{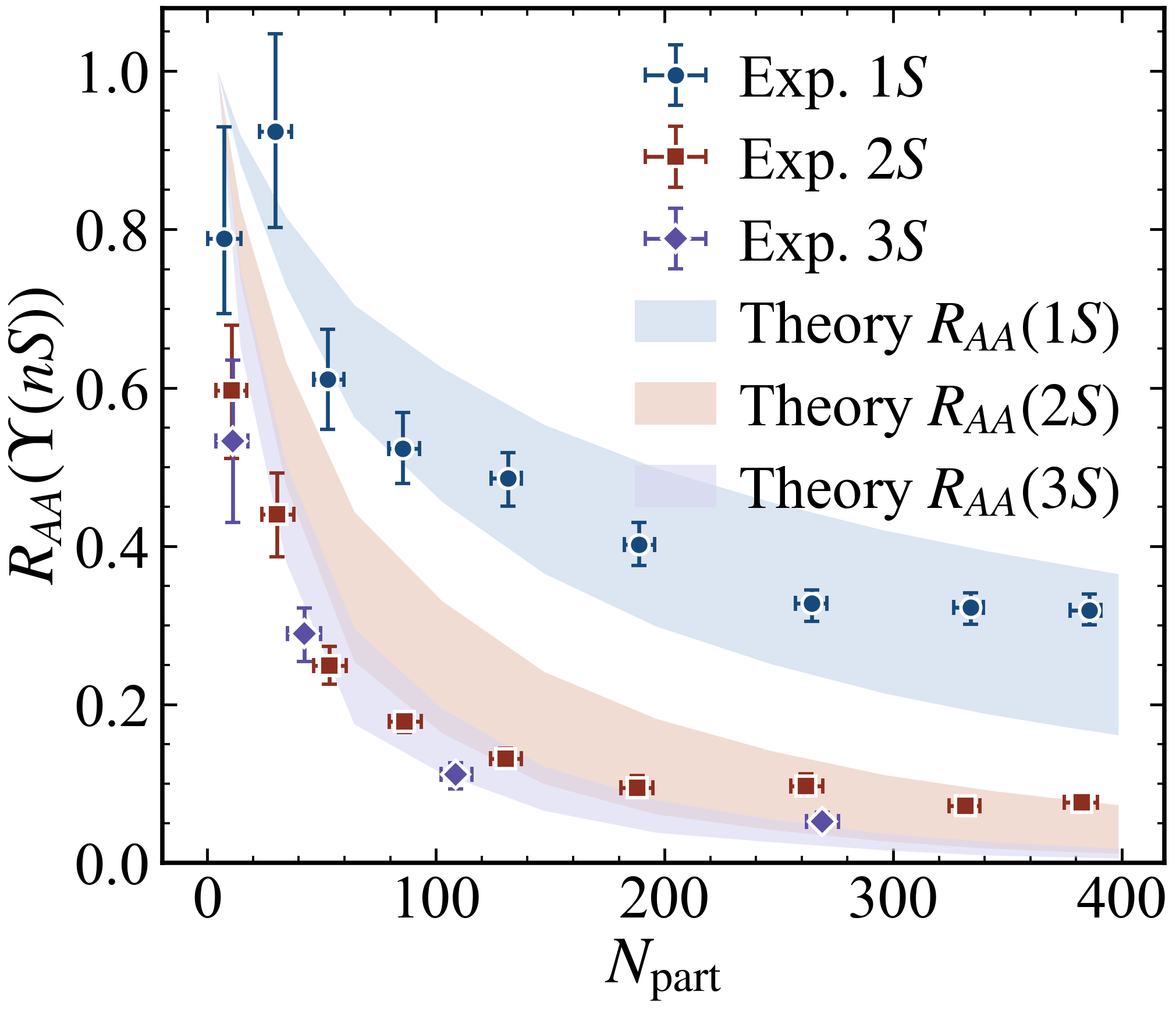}
    \caption{
    The prompt nuclear modification factors $R_{AA}$ of bottomonium $\Upsilon(1S),\Upsilon(2S),\Upsilon(3S)$ as a function of the number of participants $N_p$ in $\sqrt{s_{NN}}=5.02$ TeV Pb-Pb collisions. The bands of $R_{AA}$ originate from the uncertainty of the $V_I(T,r)$ given before. Experimental data points are cited from~\cite{CMS:2018zza,CMS:2023gfb}. \tred{Cold-nuclear-matter effects are not included in the calculation.}
    }
    \label{fig:bottomonium_RAA_AA}
\end{figure}

The Pb--Pb reference calculation employs the same vacuum Cornell real part, as used in the \(p\)--Pb calculation. The medium-induced system-size dependence is therefore mainly generated by the temperature-dependent imaginary part \(V_I(T,r)\) and by the different hydrodynamic temperature histories. In Pb-Pb collisions, while charmonium final production is dominated by the coalescence of charm and anti-charm quarks from different pairs generated in the initial parton hard scatterings, bottomonium final production, especially the ground states, are mainly from the primordial production. Therefore, the Schr\"odinger equation model remains suitable for describing bottomonium physical observables in Pb-Pb collisions at $\sqrt{s_{NN}}=5.02$ TeV. When bottomonium states propagate in the hot deconfined medium, the temperature profiles utilized in the Hamiltonian of the $b\bar b$ dipole are given by the iEBE-VISHNU package~\cite{Shen:2014vra} with a smooth initial entropy density. The detailed studies about bottomonium evolution in RHIC Au-Au and LHC Pb-Pb collisions have been investigated in~\cite{Wen:2022yjx,Liu:2026uav}. Following the same setup about the bulk medium and bottomonium initial conditions, the nuclear modification factors $R_{AA}$ of various bottomonium states $\Upsilon(1S,2S,3S)$ are given in Fig.~\ref{fig:bottomonium_RAA_AA}. The temperatures and lifetimes of the bulk medium become larger, resulting in stronger suppression over bottomonium. Furthermore, there is a clear sequential suppression pattern in the nuclear modification factors $R_{AA}$ of $(\Upsilon(1S), \Upsilon(2S), \Upsilon(3S))$, where excited bottomonium states with larger spatial size suffer stronger dissociation attributed to the radius dependence in the imaginary potential $V_I(T,r)$. The bands of $R_{AA}$ correspond to the lower and upper limits of the $V_I$ in Fig.~\ref{fig:VI}. \tred{The absolute \(R_{AA}\) results shown in Fig.~\ref{fig:bottomonium_RAA_AA} include the hot-medium suppression generated by the complex potential and the feed-down contributions, but do not include cold-nuclear-matter effects such as nuclear shadowing or Cronin broadening.}

\begin{figure}[htp]
    \centering
\includegraphics[width=\linewidth]{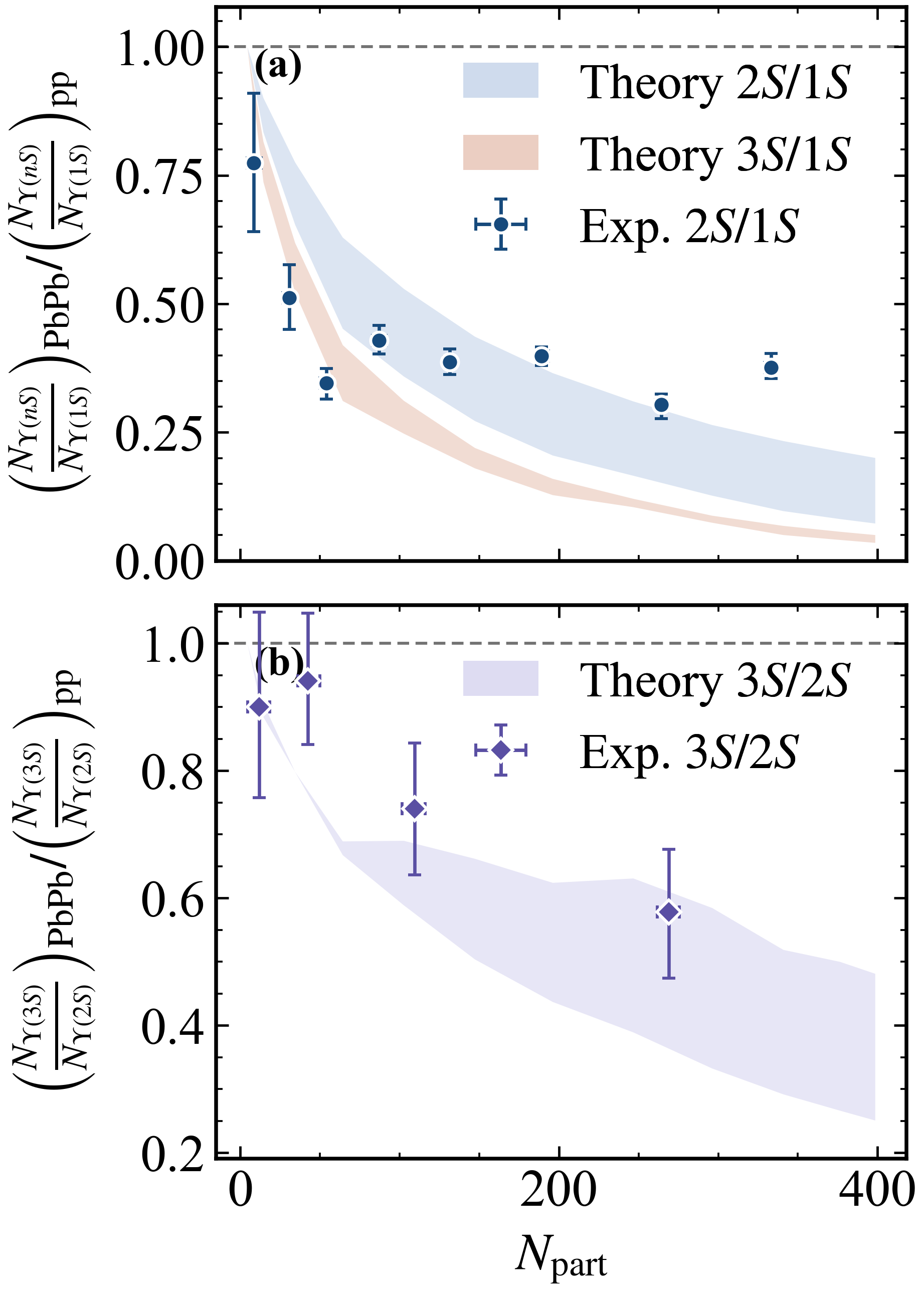}
    \caption{
    The ratio of bottomonium nuclear modification factors ($R_{AA}(\Upsilon(3S))/R_{AA}(\Upsilon(1S))$ and $R_{AA}(\Upsilon(2S))/R_{AA}(\Upsilon(1S))$ in upper panel, $R_{AA}(\Upsilon(3S))/R_{AA}(\Upsilon(2S))$ in lower panel), as a function of $N_{\rm part}$ in $\sqrt{s_{NN}}=5.02$ TeV Pb-Pb collisions. The experimental data points are cited from~~\cite{CMS:2018zza,CMS:2023gfb}.
    }
\label{fig:bottomonium_ratio2S1S_AA}
\end{figure}

With the nuclear modification factors of different bottomonium states, we also present their ratio such as $R_{AA}(\Upsilon(2S))/R_{AA}(\Upsilon(1S))$ in the upper panel of Fig.~\ref{fig:bottomonium_ratio2S1S_AA}. Theoretical bands originate from the uncertainty of the $V_I(T,r)$ in Fig.~\ref{fig:VI}. \tred{For each boundary of the theoretical band, the numerator and denominator are evaluated using the same limiting parameter set of \(V_I(T,r)\), thereby retaining their correlated response to the imaginary potential.} The theoretical calculation of $R_{AA}(\Upsilon(2S))/R_{AA}(\Upsilon(1S))$ qualitatively {describes} the experimental data in the figure. In the most central collisions, the theoretical bands continue decreasing with $N_{\rm part}$ and become smaller than the experimental data points. This may be attributed to the absence of the bottomonium regeneration in our Schr\"odinger equation model, especially for the bottomonium $\Upsilon(2S)$ and $\Upsilon(3S)$. As in most central collisions, bottom quark pair number in Pb-Pb collisions increases significantly with the number of binary collisions and results in an evident enhancement of bottomonium regeneration. This additional contribution will enhance the ratio $R_{AA}(\Upsilon(2S))/R_{AA}(\Upsilon(1S))$ and $R_{AA}(\Upsilon(3S))/R_{AA}(\Upsilon(1S))$. \ttred{For an estimate of the possible magnitude, existing bottomonium transport calculations indicate that the regeneration contribution is typically of order $R_{AA}^{\rm regen}\simeq0.05\!-\!0.10$ for $\Upsilon(1S)$ and $\Upsilon(2S)$ at $N_{\rm part}\gtrsim100$, while the corresponding absolute contribution for $\Upsilon(3S)$ is smaller~\cite{Du:2017qkv}. Incorporating the regeneration component increases the
$\Upsilon(2S)/\Upsilon(1S)$ double ratio from approximately
$0.20$ to $0.33\!-\!0.43$, bringing our theoretical calculations into better agreement with the experimental data for central collisions in the upper panel of Fig. \ref{fig:bottomonium_ratio2S1S_AA}.} In the lower panel of the  Fig.~\ref{fig:bottomonium_ratio2S1S_AA}, the ratio $R_{AA}(\Upsilon(3S))/R_{AA}(\Upsilon(2S))$ is also calculated and compared with the experimental data. Different from the bottomonium 3S/1S and 2S/1S case, the ratio $R_{AA}(\Upsilon(3S))/R_{AA}(\Upsilon(2S))$ shows nontrivial dependence on the imaginary part $V_I$ of the heavy quark potential. The behavior of the theoretical bands in the lower panel of Fig.~\ref{fig:bottomonium_ratio2S1S_AA} depends on the radius and temperature dependence in $V_I(T,r)$ and also the detailed inner evolution of $b\bar b$ wave packet with the complex heavy quark potential.

In summary, we have investigated the activity-dependent suppression of the yield ratios of heavy quarkonium in $p$-Pb collisions at $\sqrt{s_{NN}}=8.16$ TeV using a real-time Schr\"odinger evolution framework with an in-medium complex heavy-quark potential. By fixing the real part to the vacuum Cornell form and encoding final-state dissociation mechanisms in the temperature-dependent imaginary part, this approach successfully establishes a unified description across both bottomonium and charmonium flavors. Our model effectively reproduces the decreasing trends of the relative yield ratios, $\Upsilon(2S)/\Upsilon(1S)$, $\Upsilon(3S)/\Upsilon(1S)$ and $\psi(2S)/J/\psi$, as functions of charged-particle multiplicity. The observed sequential suppression pattern, where the more weakly bound and spatially extended $\Upsilon(3S)$ state exhibits stronger depletion than $\Upsilon(2S)$, is naturally explained by its higher sensitivity to the accumulated imaginary potential along the hot medium trajectories. Combined with a qualitative reference calculation for Pb-Pb collisions, which demonstrates a smoothly amplified dissociation effect in a hotter and longer-lived fireball, these results strongly support a final-state dissipative interpretation and the formation of a transient, hot deconfined QCD medium in high-multiplicity small systems.

\vspace{2cm}
\par
\medskip
\noindent\hspace{\parindent}\textbf{Acknowledgments:}\enspace
We are grateful to Shirsendu Nanda, Shuhan Zheng, Shuzhe Shi, and Xiaojian Du for helpful discussions and valuable comments.
This work is supported by the
National Natural Science Foundation of China (NSFC)
under Grant Nos. 12575149 and 12175165.
\par

\bibliographystyle{apsrev4-1}
\bibliography{ref_pA}

@article{Liu:2026uav,
    author = "Liu, Jiamin and Zhou, Kai and Chen, Baoyi",
    title = "{Unified Extraction of In-Medium Heavy Quark Potentials from RHIC to LHC Energies via Deep Learning}",
    eprint = "2604.09198",
    archivePrefix = "arXiv",
    primaryClass = "nucl-th",
    journal = {arXiv e-prints},
    month = "4",
    year = "2026"
}

@article{CMS:2020fae,
    author = "Sirunyan, Albert M and others",
    collaboration = "CMS",
    title = "{Investigation into the event-activity dependence of $\Upsilon$(nS) relative production in proton-proton collisions at $ \sqrt{s} $ = 7 TeV}",
    eprint = "2007.04277",
    archivePrefix = "arXiv",
    primaryClass = "hep-ex",
    reportNumber = "CMS-BPH-14-009, CERN-EP-2020-075",
    doi = "10.1007/JHEP11(2020)001",
    journal = "JHEP",
    volume = "11",
    pages = "001",
    year = "2020"
}

@article{ParticleDataGroup:2018ovx,
    author = "Tanabashi, M. and others",
    collaboration = "Particle Data Group",
    title = "{Review of Particle Physics}",
    doi = "10.1103/PhysRevD.98.030001",
    journal = "Phys. Rev. D",
    volume = "98",
    number = "3",
    pages = "030001",
    year = "2018"
}

@article{Satz:2005hx,
    author = "Satz, Helmut",
    title = "{Colour deconfinement and quarkonium binding}",
    eprint = "hep-ph/0512217",
    archivePrefix = "arXiv",
    doi = "10.1088/0954-3899/32/3/R01",
    journal = "J. Phys. G",
    volume = "32",
    pages = "R25",
    year = "2006"
}

@article{Wen:2022utn,
    author = "Wen, Liuyuan and Du, Xiaojian and Shi, Shuzhe and Chen, Baoyi",
    title = "{Investigating color screening in proton-nucleus collisions with complex potentials*}",
    eprint = "2205.07520",
    archivePrefix = "arXiv",
    primaryClass = "nucl-th",
    doi = "10.1088/1674-1137/ac7fe6",
    journal = "Chin. Phys. C",
    volume = "46",
    number = "11",
    pages = "114102",
    year = "2022"
}

@article{Matsui:1986dk,
    author = "Matsui, T. and Satz, H.",
    title = "{J/psi Suppression by Quark-Gluon Plasma Formation}",
    reportNumber = "BNL-38344",
    doi = "10.1016/0370-2693(86)91404-8",
    journal = "Phys. Lett. B",
    volume = "178",
    pages = "416--422",
    year = "1986"
}

@article{Brambilla:2010cs,
    author = "Brambilla, N. and others",
    title = "{Heavy quarkonium: progress, puzzles, and opportunities}",
    eprint = "1010.5827",
    archivePrefix = "arXiv",
    primaryClass = "hep-ph",
    doi = "10.1140/epjc/s10052-010-1534-9",
    journal = "Eur. Phys. J. C",
    volume = "71",
    pages = "1534",
    year = "2011"
}

@article{Andronic:2015wma,
    author = "Andronic, A. and others",
    title = "{Heavy-flavour and quarkonium production in the LHC era: from proton-proton to heavy-ion collisions}",
    eprint = "1506.03981",
    archivePrefix = "arXiv",
    primaryClass = "nucl-ex",
    doi = "10.1140/epjc/s10052-015-3819-5",
    journal = "Eur. Phys. J. C",
    volume = "76",
    number = "3",
    pages = "107",
    year = "2016"
}

@article{Rapp:2008tf,
    author = "Rapp, R. and Blaschke, D. and Crochet, P.",
    title = "{Charmonium and bottomonium production in heavy-ion collisions}",
    eprint = "0807.2470",
    archivePrefix = "arXiv",
    primaryClass = "hep-ph",
    doi = "10.1016/j.ppnp.2008.12.002",
    journal = "Prog. Part. Nucl. Phys.",
    volume = "65",
    pages = "209--266",
    year = "2010"
}

@article{Emerick:2011xu,
    author = "Emerick, Anthony and Zhao, Xingbo and Rapp, Ralf",
    title = "{Bottomonia in the Quark-Gluon Plasma and their Production at RHIC and LHC}",
    eprint = "1111.6537",
    archivePrefix = "arXiv",
    primaryClass = "hep-ph",
    doi = "10.1140/epja/i2012-12072-y",
    journal = "Eur. Phys. J. A",
    volume = "48",
    pages = "72",
    year = "2012"
}

@article{Du:2017qkv,
    author = "Du, Xiaojian and He, Min and Rapp, Ralf",
    title = "{Color Screening and Regeneration of Bottomonia in High-Energy Heavy-Ion Collisions}",
    eprint = "1706.08670",
    archivePrefix = "arXiv",
    primaryClass = "hep-ph",
    doi = "10.1103/PhysRevC.96.054901",
    journal = "Phys. Rev. C",
    volume = "96",
    number = "5",
    pages = "054901",
    year = "2017"
}

@article{Laine:2006ns,
    author = "Laine, M. and Philipsen, O. and Romatschke, P. and Tassler, M.",
    title = "{Real-time static potential in hot QCD}",
    eprint = "hep-ph/0611300",
    archivePrefix = "arXiv",
    doi = "10.1088/1126-6708/2007/03/054",
    journal = "JHEP",
    volume = "03",
    pages = "054",
    year = "2007"
}

@article{Burnier:2007qm,
    author = "Laine, M. and Philipsen, O. and Tassler, M.",
    title = "{Thermal imaginary part of a real-time static potential from classical lattice gauge theory simulations}",
    eprint = "0707.2458",
    archivePrefix = "arXiv",
    primaryClass = "hep-lat",
    doi = "10.1088/1126-6708/2007/09/066",
    journal = "JHEP",
    volume = "09",
    pages = "066",
    year = "2007"
}

@article{Akamatsu:2020ypb,
    author = "Akamatsu, Yukinao",
    title = "{Quarkonium in Quark-Gluon Plasma: Open Quantum System Approaches Re-examined}",
    eprint = "2009.10559",
    archivePrefix = "arXiv",
    primaryClass = "nucl-th",
    doi = "10.1016/j.ppnp.2021.103932",
    journal = "Prog. Part. Nucl. Phys.",
    volume = "123",
    pages = "103932",
    year = "2022"
}

@article{Digal:2001ue,
    author = "Digal, S. and Petreczky, P. and Satz, H.",
    title = "{Sequential quarkonium suppression}",
    eprint = "hep-ph/0110406",
    archivePrefix = "arXiv",
    doi = "10.1103/PhysRevD.64.094015",
    journal = "Phys. Rev. D",
    volume = "64",
    pages = "094015",
    year = "2001"
}

@article{Strickland:2011aa,
    author = "Strickland, Michael and Bazow, Dennis",
    title = "{Thermal Bottomonium Suppression at RHIC and LHC}",
    eprint = "1112.2761",
    archivePrefix = "arXiv",
    primaryClass = "nucl-th",
    doi = "10.1016/j.nuclphysa.2012.02.003",
    journal = "Nucl. Phys. A",
    volume = "879",
    pages = "25--58",
    year = "2012"
}

@article{Margotta:2011ta,
    author = "Margotta, Matthew and McCarty, Kyle and McGahan, Christina and Strickland, Michael and Yager-Elorriaga, David",
    title = "{Quarkonium states in a complex-valued potential}",
    eprint = "1101.4651",
    archivePrefix = "arXiv",
    primaryClass = "hep-ph",
    doi = "10.1103/PhysRevD.84.069902",
    journal = "Phys. Rev. D",
    volume = "83",
    pages = "105019",
    year = "2011",
    note = "[Erratum: Phys.Rev.D 84, 069902 (2011)]"
}

@article{Krouppa:2017jlg,
    author = "Krouppa, Brandon and Rothkopf, Alexander and Strickland, Michael",
    title = "{Bottomonium suppression using a lattice QCD vetted potential}",
    eprint = "1710.02319",
    archivePrefix = "arXiv",
    primaryClass = "hep-ph",
    doi = "10.1103/PhysRevD.97.016017",
    journal = "Phys. Rev. D",
    volume = "97",
    number = "1",
    pages = "016017",
    year = "2018"
}

@article{Wen:2022yjx,
    author = "Wen, Liuyuan and Chen, Baoyi",
    title = "{Bottomonium sequential suppression and strong heavy-quark potential in heavy-ion collisions}",
    eprint = "2208.10050",
    archivePrefix = "arXiv",
    primaryClass = "nucl-th",
    doi = "10.1016/j.physletb.2023.137774",
    journal = "Phys. Lett. B",
    volume = "839",
    pages = "137774",
    year = "2023"
}

@article{CMS:2013dla,
    author = "Chatrchyan, Serguei and others",
    collaboration = "CMS",
    title = "{Event activity dependence of \(\Upsilon(nS)\) production in \(\sqrt{s_{NN}}=5.02\) TeV pPb and \(\sqrt{s}=2.76\) TeV pp collisions}",
    eprint = "1312.6300",
    archivePrefix = "arXiv",
    primaryClass = "nucl-ex",
    reportNumber = "CMS-HIN-13-003, CERN-PH-EP-2013-219",
    doi = "10.1007/JHEP04(2014)103",
    journal = "JHEP",
    volume = "04",
    pages = "103",
    year = "2014"
}

@article{CMS:2023gfb,
    author = "Tumasyan, Armen and others",
    collaboration = "CMS",
    title = "{Observation of the \(\Upsilon(3S)\) Meson and Suppression of \(\Upsilon\) States in Pb-Pb Collisions at \(\sqrt{s_{NN}}=5.02\) TeV}",
    eprint = "2303.17026",
    archivePrefix = "arXiv",
    primaryClass = "hep-ex",
    reportNumber = "CMS-HIN-21-007, CERN-EP-2023-011",
    doi = "10.1103/PhysRevLett.133.022302",
    journal = "Phys. Rev. Lett.",
    volume = "133",
    number = "2",
    pages = "022302",
    year = "2024"
}

@article{CMS:2025psi2SJpsi_pPb,
    author = "Chekhovsky, V. and others",
    collaboration = "CMS",
    title = "{Observation of the Charged-Particle Multiplicity Dependence of \(\sigma_{\psi(2S)}/\sigma_{J/\psi}\) in p-Pb Collisions at 8.16 TeV}",
    eprint = "2503.02139",
    archivePrefix = "arXiv",
    primaryClass = "hep-ex",
    reportNumber = "CMS-HIN-24-001, CERN-EP-2024-332",
    doi = "10.1103/c9wp-5tq3",
    journal = "Phys. Rev. Lett.",
    volume = "135",
    number = "9",
    pages = "092301",
    year = "2025"
}

@article{ALICE:2014wnc,
    author = "Abelev, Betty Bezverkhny and others",
    collaboration = "ALICE",
    title = "{Suppression of \(\Upsilon(1S)\) at forward rapidity in Pb-Pb collisions at \(\sqrt{s_{NN}}=2.76\) TeV}",
    eprint = "1405.4493",
    archivePrefix = "arXiv",
    primaryClass = "nucl-ex",
    doi = "10.1016/j.physletb.2014.10.001",
    journal = "Phys. Lett. B",
    volume = "738",
    pages = "361--372",
    year = "2014"
}

@article{CMS:2010ifv,
    author = "Khachatryan, Vardan and others",
    collaboration = "CMS",
    title = "{Observation of Long-Range Near-Side Angular Correlations in Proton-Proton Collisions at the LHC}",
    eprint = "1009.4122",
    archivePrefix = "arXiv",
    primaryClass = "hep-ex",
    reportNumber = "CMS-QCD-10-002, CERN-PH-EP-2010-031",
    doi = "10.1007/JHEP09(2010)091",
    journal = "JHEP",
    volume = "09",
    pages = "091",
    year = "2010"
}

@article{CMS:2012qk,
    author = "Chatrchyan, Serguei and others",
    collaboration = "CMS",
    title = "{Observation of Long-Range Near-Side Angular Correlations in Proton-Lead Collisions at the LHC}",
    eprint = "1210.5482",
    archivePrefix = "arXiv",
    primaryClass = "nucl-ex",
    reportNumber = "CMS-HIN-12-005, CERN-PH-EP-2012-320, CMS-HIN-12-015",
    doi = "10.1016/j.physletb.2012.11.025",
    journal = "Phys. Lett. B",
    volume = "718",
    pages = "795--814",
    year = "2013"
}

@article{ALICE:2012eyl,
    author = "Abelev, Betty Bezverkhny and others",
    collaboration = "ALICE",
    title = "{Long-range angular correlations on the near and away side in p-Pb collisions at \(\sqrt{s_{NN}}=5.02\) TeV}",
    eprint = "1212.2001",
    archivePrefix = "arXiv",
    primaryClass = "nucl-ex",
    reportNumber = "CERN-PH-EP-2012-359",
    doi = "10.1016/j.physletb.2013.01.012",
    journal = "Phys. Lett. B",
    volume = "719",
    pages = "29--41",
    year = "2013"
}

@article{Nagle:2018nvi,
    author = "Nagle, James L. and Zajc, William A.",
    title = "{Small System Collectivity in Relativistic Hadronic and Nuclear Collisions}",
    eprint = "1801.03477",
    archivePrefix = "arXiv",
    primaryClass = "nucl-ex",
    doi = "10.1146/annurev-nucl-101916-123209",
    journal = "Ann. Rev. Nucl. Part. Sci.",
    volume = "68",
    pages = "211--235",
    year = "2018"
}

@article{Ferreiro:2014bia,
    author = "Ferreiro, E. G.",
    title = "{Excited charmonium suppression in proton--nucleus collisions as a consequence of comovers}",
    eprint = "1411.0549",
    archivePrefix = "arXiv",
    primaryClass = "hep-ph",
    doi = "10.1016/j.physletb.2015.07.066",
    journal = "Phys. Lett. B",
    volume = "749",
    pages = "98--103",
    year = "2015"
}

@article{Arleo:2012rs,
    author = "Arleo, Francois and Peigne, Stephane",
    title = "{J/psi suppression in p-A collisions from parton energy loss in cold QCD matter}",
    eprint = "1204.4609",
    archivePrefix = "arXiv",
    primaryClass = "hep-ph",
    doi = "10.1103/PhysRevLett.109.122301",
    journal = "Phys. Rev. Lett.",
    volume = "109",
    pages = "122301",
    year = "2012"
}

@article{Shen:2014vra,
    author = "Shen, Chun and Qiu, Zhi and Song, Huichao and Bernhard, Jonah and Bass, Steffen and Heinz, Ulrich",
    title = "{The iEBE-VISHNU code package for relativistic heavy-ion collisions}",
    eprint = "1409.8164",
    archivePrefix = "arXiv",
    primaryClass = "nucl-th",
    doi = "10.1016/j.cpc.2015.08.039",
    journal = "Comput. Phys. Commun.",
    volume = "199",
    pages = "61--85",
    year = "2016"
}

@article{Du:2018wsj,
    author = "Du, Xiaojian and Rapp, Ralf",
    title = "{In-Medium Charmonium Production in Proton-Nucleus Collisions}",
    eprint = "1808.10014",
    archivePrefix = "arXiv",
    primaryClass = "nucl-th",
    doi = "10.1007/JHEP03(2019)015",
    journal = "JHEP",
    volume = "03",
    pages = "015",
    year = "2019"
}

@article{LHCb:2018psc,
    author = "Aaij, Roel and others",
    collaboration = "LHCb",
    title = "{Study of $\Upsilon$ production in $p$Pb collisions at $\sqrt{s_{NN}}=8.16$ TeV}",
    eprint = "1810.07655",
    archivePrefix = "arXiv",
    primaryClass = "hep-ex",
    reportNumber = "LHCb-PAPER-2018-035, CERN-EP-2018-267",
    doi = "10.1007/JHEP11(2018)194",
    journal = "JHEP",
    volume = "11",
    pages = "194",
    year = "2018",
    note = "[Erratum: JHEP 02, 093 (2020)]"
}

@article{CMS:2018zza,
    author = "Sirunyan, Albert M and others",
    collaboration = "CMS",
    title = "{Measurement of nuclear modification factors of $\Upsilon$(1S), $\Upsilon$(2S), and $\Upsilon$(3S) mesons in PbPb collisions at $\sqrt{s_{_\mathrm{NN}}} =$ 5.02 TeV}",
    eprint = "1805.09215",
    archivePrefix = "arXiv",
    primaryClass = "hep-ex",
    reportNumber = "CMS-HIN-16-023, CERN-EP-2018-110",
    doi = "10.1016/j.physletb.2019.01.006",
    journal = "Phys. Lett. B",
    volume = "790",
    pages = "270--293",
    year = "2019"
}

@article{Brambilla:2020qwo,
    author = "Brambilla, Nora and Escobedo, Miguel {\'A}ngel and Strickland, Michael and Vairo, Antonio and Vander Griend, Peter and Weber, Johannes Heinrich",
    title = "{Bottomonium suppression in an open quantum system using the quantum trajectories method}",
    eprint = "2012.01240",
    archivePrefix = "arXiv",
    primaryClass = "hep-ph",
    reportNumber = "TUM-EFT-140-20, HU-EP-20-36-RTG",
    doi = "10.1007/JHEP05(2021)136",
    journal = "JHEP",
    volume = "05",
    pages = "136",
    year = "2021"
}

@article{Brambilla:2022ynh,
    author = "Brambilla, Nora and Escobedo, Miguel Angel and Islam, Ajaharul and Strickland, Michael and Tiwari, Anurag and Vairo, Antonio and Vander Griend, Peter",
    title = "{Heavy quarkonium dynamics at next-to-leading order in the binding energy over temperature}",
    eprint = "2205.10289",
    archivePrefix = "arXiv",
    primaryClass = "hep-ph",
    journal = "JHEP",
    volume = "08",
    pages = "303",
    year = "2022",
    doi = "10.1007/JHEP08(2022)303"
}

@article{Wei:2026iwq,
    author = "Wei, Linyuan and Hu, Jin and Huang, Anping and Liu, Yunpeng and Chen, Baoyi",
    title = "{Quantum simulation of bottomonium dynamics in the quark-gluon plasma via the Lindblad equation}",
    eprint = "2608.06754",
    archivePrefix = "arXiv",
    primaryClass = "nucl-th",
    month = "8",
    year = "2026"
}

@article{Liu:2012tn,
    author = "Liu, Yunpeng and Xu, Nu and Zhuang, Pengfei",
    title = "{Velocity Dependence of Charmonium Dissociation Temperature in High-Energy Nuclear Collisions}",
    eprint = "1210.7449",
    archivePrefix = "arXiv",
    primaryClass = "nucl-th",
    journal = "Phys. Lett. B",
    volume = "724",
    pages = "73--76",
    year = "2013"
}

@article{CMS:2013qur,
    author = "Chatrchyan, Serguei and others",
    collaboration = "CMS",
    title = "{Measurement of the $\Upsilon(1S), \Upsilon(2S)$, and $\Upsilon(3S)$ Cross Sections in $pp$ Collisions at $\sqrt{s}$ = 7 TeV}",
    eprint = "1303.5900",
    archivePrefix = "arXiv",
    primaryClass = "hep-ex",
    reportNumber = "CMS-BPH-11-001, CERN-PH-EP-2012-373",
    doi = "10.1016/j.physletb.2013.10.033",
    journal = "Phys. Lett. B",
    volume = "727",
    pages = "101--125",
    year = "2013"
}

@article{Chen:2018kfo,
    author = "Chen, Baoyi",
    title = "{Thermal production of charmonia in Pb-Pb collisions at $\sqrt {s_{NN}} = $ 5.02 TeV}",
    eprint = "1811.11393",
    archivePrefix = "arXiv",
    primaryClass = "nucl-th",
    doi = "10.1088/1674-1137/43/12/124101",
    journal = "Chin. Phys. C",
    volume = "43",
    number = "12",
    pages = "124101",
    year = "2019"
}

@article{Zhao:2021voa,
    author = "Zhao, Jiaxing and Chen, Baoyi and Zhuang, Pengfei",
    title = "{Charmonium triangular flow in high energy nuclear collisions}",
    eprint = "2112.00293",
    archivePrefix = "arXiv",
    primaryClass = "hep-ph",
    doi = "10.1103/PhysRevC.105.034902",
    journal = "Phys. Rev. C",
    volume = "105",
    number = "3",
    pages = "034902",
    year = "2022"
}

@article{Zhou:2014kka,
    author = "Zhou, Kai and Xu, Nu and Xu, Zhe and Zhuang, Pengfei",
    title = "{Medium effects on charmonium production at ultrarelativistic energies available at the CERN Large Hadron Collider}",
    eprint = "1401.5845",
    archivePrefix = "arXiv",
    primaryClass = "nucl-th",
    doi = "10.1103/PhysRevC.89.054911",
    journal = "Phys. Rev. C",
    volume = "89",
    number = "5",
    pages = "054911",
    year = "2014"
}

@article{Singh:2021evv,
    author = "Singh, Captain R. and Deb, Suman and Sahoo, Raghunath and Alam, Jan-e",
    title = "{Charmonium suppression in ultra-relativistic proton{\textendash}proton collisions at LHC energies: a hint for QGP in small systems}",
    eprint = "2109.07967",
    archivePrefix = "arXiv",
    primaryClass = "hep-ph",
    doi = "10.1140/epjc/s10052-022-10500-z",
    journal = "Eur. Phys. J. C",
    volume = "82",
    number = "6",
    pages = "542",
    year = "2022"
}

@article{Yan:2006ve,
    author = "Yan, Li and Zhuang, Pengfei and Xu, Nu",
    title = "{Competition between J / psi suppression and regeneration in quark-gluon plasma}",
    eprint = "nucl-th/0608010",
    archivePrefix = "arXiv",
    doi = "10.1103/PhysRevLett.97.232301",
    journal = "Phys. Rev. Lett.",
    volume = "97",
    pages = "232301",
    year = "2006"
}

@article{Chen:2016dke,
    author = "Chen, Baoyi and Guo, Tiecheng and Liu, Yunpeng and Zhuang, Pengfei",
    title = "{Cold and Hot Nuclear Matter Effects on Charmonium Production in p+Pb Collisions at LHC Energy}",
    eprint = "1607.07927",
    archivePrefix = "arXiv",
    primaryClass = "nucl-th",
    doi = "10.1016/j.physletb.2016.12.021",
    journal = "Phys. Lett. B",
    volume = "765",
    pages = "323--327",
    year = "2017"
}

@article{Chen:2015iga,
    author = "Chen, Baoyi",
    title = "{Detailed rapidity dependence of $J/\psi$ production at energies available at the Large Hadron Collider}",
    eprint = "1510.07466",
    archivePrefix = "arXiv",
    primaryClass = "hep-ph",
    doi = "10.1103/PhysRevC.93.054905",
    journal = "Phys. Rev. C",
    volume = "93",
    number = "5",
    pages = "054905",
    year = "2016"
}

@article{Zhao:2017yan,
    author = "Zhao, Jiaxing and Chen, Baoyi",
    title = "{Strong diffusion effect of charm quarks on J /$\psi$ production in Pb{\textendash}Pb collisions at the LHC}",
    eprint = "1705.04558",
    archivePrefix = "arXiv",
    primaryClass = "nucl-th",
    doi = "10.1016/j.physletb.2017.11.014",
    journal = "Phys. Lett. B",
    volume = "776",
    pages = "17--21",
    year = "2018"
}

@article{Andronic:2003zv,
    author = "Andronic, A. and Braun-Munzinger, P. and Redlich, K. and Stachel, J.",
    title = "{Statistical hadronization of charm in heavy ion collisions at SPS, RHIC and LHC}",
    eprint = "nucl-th/0303036",
    archivePrefix = "arXiv",
    doi = "10.1016/j.physletb.2003.07.066",
    journal = "Phys. Lett. B",
    volume = "571",
    pages = "36--44",
    year = "2003"
}

@article{Rothkopf:2011db,
    author = "Rothkopf, Alexander and Hatsuda, Tetsuo and Sasaki, Shoichi",
    title = "{Complex Heavy-Quark Potential at Finite Temperature from Lattice QCD}",
    eprint = "1108.1579",
    archivePrefix = "arXiv",
    primaryClass = "hep-lat",
    reportNumber = "TKYNT-11-10, RIKEN-QHP-8, BI-TP-2011-23",
    doi = "10.1103/PhysRevLett.108.162001",
    journal = "Phys. Rev. Lett.",
    volume = "108",
    pages = "162001",
    year = "2012"
}

@article{Brambilla:2016wgg,
    author = "Brambilla, Nora and Escobedo, Miguel A. and Soto, Joan and Vairo, Antonio",
    title = "{Quarkonium suppression in heavy-ion collisions: an open quantum system approach}",
    eprint = "1612.07248",
    archivePrefix = "arXiv",
    primaryClass = "hep-ph",
    reportNumber = "ICCUB-16-044, TUM-EFT-55-14",
    doi = "10.1103/PhysRevD.96.034021",
    journal = "Phys. Rev. D",
    volume = "96",
    number = "3",
    pages = "034021",
    year = "2017"
}

@article{Zhao:2017rgg,
    author = "Zhao, Wenbin and Zhou, You and Xu, Haojie and Deng, Weitian and Song, Huichao",
    title = "{Hydrodynamic collectivity in proton{\textendash}proton collisions at 13 TeV}",
    eprint = "1801.00271",
    archivePrefix = "arXiv",
    primaryClass = "nucl-th",
    doi = "10.1016/j.physletb.2018.03.022",
    journal = "Phys. Lett. B",
    volume = "780",
    pages = "495--500",
    year = "2018"
}

@article{Schenke:2010nt,
author = "Schenke, Bjoern and Jeon, Sangyong and Gale, Charles",
title = "{(3+1)D hydrodynamic simulation of relativistic heavy-ion collisions}",
eprint = "1004.1408",
archivePrefix = "arXiv",
primaryClass = "hep-ph",
doi = "10.1103/PhysRevC.82.014903",
journal = "Phys. Rev. C",
volume = "82",
pages = "014903",
year = "2010"
}

@article{Zheng:2025bdc,
    author = "Zheng, Shuhan and Chen, Baoyi and Du, Xiaojian and Shi, Shuzhe",
    title = "{Data-Driven Analysis for the Bottomonium Potential in the Quark-Gluon Plasma}",
    eprint = "2512.11536",
    archivePrefix = "arXiv",
    primaryClass = "nucl-th",
    month = "12",
    year = "2025"
}

@article{Zhao:2020wcd,
    author = "Zhao, Wenbin and Ko, Che Ming and Liu, Yu-Xin and Qin, Guang-You and Song, Huichao",
    title = "{Probing the Partonic Degrees of Freedom in High-Multiplicity $p-Pb$ collisions at $\sqrt {s_{NN}}$ = 5.02  TeV}",
    eprint = "1911.00826",
    archivePrefix = "arXiv",
    primaryClass = "nucl-th",
    doi = "10.1103/PhysRevLett.125.072301",
    journal = "Phys. Rev. Lett.",
    volume = "125",
    number = "7",
    pages = "072301",
    year = "2020"
}

@article{Schenke:2010rr,
    author = "Schenke, Bjorn and Jeon, Sangyong and Gale, Charles",
    title = "{Elliptic and triangular flow in event-by-event (3+1)D viscous hydrodynamics}",
    eprint = "1009.3244",
    archivePrefix = "arXiv",
    primaryClass = "hep-ph",
    doi = "10.1103/PhysRevLett.106.042301",
    journal = "Phys. Rev. Lett.",
    volume = "106",
    pages = "042301",
    year = "2011"
}

@article{Eskola:2016oht,
    author = "Eskola, Kari J. and Paakkinen, Petja and Paukkunen, Hannu and Salgado, Carlos A.",
    title = "{EPPS16: Nuclear parton distributions with LHC data}",
    eprint = "1612.05741",
    archivePrefix = "arXiv",
    primaryClass = "hep-ph",
    doi = "10.1140/epjc/s10052-017-4725-9",
    journal = "Eur. Phys. J. C",
    volume = "77",
    number = "3",
    pages = "163",
    year = "2017"
}

@article{Cronin:1974zm,
    author = "Cronin, J. W. and Frisch, Henry J. and Shochet, M. J. and Boymond, J. P. and Mermod, R. and Piroue, P. A. and Sumner, Richard L.",
    editor = "Smith, J. R.",
    collaboration = "E100",
    title = "{Production of hadrons with large transverse momentum at 200, 300, and 400 GeV}",
    reportNumber = "PRINT-74-1181 (EFI,CHICAGO), FERMILAB-CONF-74-148-E",
    doi = "10.1103/PhysRevD.11.3105",
    journal = "Phys. Rev. D",
    volume = "11",
    pages = "3105--3123",
    year = "1975"
}

@article{Burnier:2014ssa,
    author = "Burnier, Yannis and Kaczmarek, Olaf and Rothkopf, Alexander",
    title = "{Static quark-antiquark potential in the quark-gluon plasma from lattice QCD}",
    eprint = "1410.2546",
    archivePrefix = "arXiv",
    primaryClass = "hep-lat",
    reportNumber = "BI-TP-2014-21",
    doi = "10.1103/PhysRevLett.114.082001",
    journal = "Phys. Rev. Lett.",
    volume = "114",
    number = "8",
    pages = "082001",
    year = "2015"
}

@article{Burnier:2016mxc,
    author = "Burnier, Yannis and Rothkopf, Alexander",
    title = "{Complex heavy-quark potential and Debye mass in a gluonic medium from lattice QCD}",
    eprint = "1607.04049",
    archivePrefix = "arXiv",
    primaryClass = "hep-lat",
    doi = "10.1103/PhysRevD.95.054511",
    journal = "Phys. Rev. D",
    volume = "95",
    number = "5",
    pages = "054511",
    year = "2017"
}

@article{LHCb:2014ngh,
    author = "Aaij, Roel and others",
    collaboration = "LHCb",
    title = "{Study of $\chi _{{\mathrm {b}}}$ meson production in $\mathrm {p} $ $\mathrm {p} $ collisions at $\sqrt{s}=7$ and $8{\mathrm {\,TeV}} $ and observation of the decay $\chi _{{\mathrm {b}}}\mathrm {(3P)} \rightarrow \Upsilon \mathrm {(3S)} {\gamma } $}",
    eprint = "1407.7734",
    archivePrefix = "arXiv",
    primaryClass = "hep-ex",
    reportNumber = "CERN-PH-EP-2014-178, LHCB-PAPER-2014-031",
    doi = "10.1140/epjc/s10052-014-3092-z",
    journal = "Eur. Phys. J. C",
    volume = "74",
    number = "10",
    pages = "3092",
    year = "2014"
}

@article{Boyd:2023ybk,
    author = "Boyd, Jacob and Thapa, Sabin and Strickland, Michael",
    title = "{Transverse momentum dependent feed-down fractions for bottomonium production}",
    eprint = "2307.03841",
    archivePrefix = "arXiv",
    primaryClass = "hep-ph",
    doi = "10.1103/PhysRevD.108.094024",
    journal = "Phys. Rev. D",
    volume = "108",
    number = "9",
    pages = "094024",
    year = "2023"
}

\end{document}